\documentstyle[12pt,a4]{article}
\begin{document}
\author{B. G\"{o}n\"{u}l, O. \"{O}zer, Y. Can\c{c}elik ,and  M. Ko\c{c}ak
\and Department of Engineering Physics, University of Gaziantep,
\and 27310 Gaziantep-T\"{u}rkiye}
\title{Hamiltonian  hierarchy and the Hulth\'{e}n potential}
\maketitle
\begin{abstract}
We deal with the Hamiltonian hierarchy problem of the Hulth\'{e}n potential 
within the frame of the supersymmetric quantum mechanics and find that 
the associated superymmetric partner potentials simulate the effect 
of the centrifugal barrier. Incorporating the supersymmetric solutions 
and using the first-order perturbation theory we obtain an expression for 
the energy levels of theHulth\'{e}n potential which gives satisfactory 
values for the non-zero angular momentum states.
\end{abstract}
{\bf Pacs Numbers}:~03.65.Fd, 03.65.Ge


Published in PLA 275 (2000) p. 238

\section{Introduction}
TheHulth\'{e}n potential [1,2] is one of the important short-range 
potentials in physics. The potential is given by
\begin{equation}
V(r)=-\frac{Ze^{2}\delta e^{-\delta r}} {(1-e^{-\delta r})}
\end{equation}
where $Z$ is a constant and $\delta$ is the screening parameter.
If the potential is used for atoms, the $Z$ is identified
with the atomic number. This potential is a special case of the
Eckart potential \cite{eckart} which has been widely used in
several branches of physics and its bound-state and scattering
properties have been investigated by a variety of techniques (see
e.g., \cite{varshni} and references therein).

Unfortunately, the radial Schr\"{o}dinger equation for the Hulth\'{e}n 
potential can be solved analytically only for the states with zero 
angular momentum [1,2,5,6]. For $\ell\neq 0$, a number of methods have 
been employed to evaluate bound-state energies numerically [4,7-17]. 
In connection with this, we present in this letter a method
within the frame of supersymmetric quantum mechanics (SUSYQM) using an 
effective Hulth\'{e}n potential for non-zero angular momentum states, 
which can be solved analytically.

In supersymmetric quantum mechanics (for a recent review see 
\cite{cooper1}) one often deals with hierarchy problem. Within the 
context of the SUSYQM one can generate a Hamiltonian hierarchy, 
the adjacent members of which are supersymmetric partners in that 
they share the same eigenvalue spectrum except for the missing 
ground state. In the case of Coulomb potential $V_{c}(r)$ , the 
Hamiltonian hierarchy corresponds to the addition of an appropriate 
centrifugal potential and the so-called accidental degenarcy is 
recovered as a natural consequence \cite{sukumar}. In this letter we
shall examine the implication of the Hamiltonian hierarchy for the 
Hulth\'{e}n potential. At small values of the radial coordinate $r$, 
the Hulth\'{e}n potential behaves like a Coulomb potential whereas 
for large values of $r$ it decreases exponentially so that its capacity 
for bound state is smaller then $V_{c}(r)$. In contrast to the 
Hulth\'{e}n potential, the Coulomb problem is analytically solvable 
for all energies and all angular momenta. Because of the similarity 
and points of contrast mentioned above between Coulomb and Hulth\'{e}n 
potentials, it may be of considerable interest to generate the 
supersymmetric partners of the Hulthén potential and study their 
eigenfunctions and eigenvalues. In the following we outline the 
basic idea of the SUSYQM and set up the formalism for dealing 
with the Hulth\'{e}n problem.

\section{ Supersymmetric solution of theHulth\'{e}n potential}
The supersymmetric formalism has already been used to study some 
aspects of the Hulth\'{e}n potential [15,17]. Here the exact 
analytical solution for this potential is re-obtained for 
$\ell=0$ state in the light of the works described  in 
Refs. [15-17,20] to show the consistency of the method and see 
how partners of the Hulth\'{e}n potential simulate the effect 
of the centrifugal barrier, which makes clear the physics 
behind partner Hamiltonians in connection with the states having
$\ell\neq0$ angular momenta.

In the approach followed here the first step taken is to 
look for an effective potential similar to the original 
Hulth\'{e}n potential. Inspired by SUSYQM, we propose
an ansatz for the superpotential,
\begin{equation}
W_{\ell+1}(r)=-\frac{\hbar}{\sqrt{2m}}\frac{(\ell+1)\delta e^{-\delta
r}}{(1-e^-\delta r)}+\sqrt{\frac{m}{2}}\frac{e^2}{\hbar}\left[
\frac{1}{(\ell+1)}-\frac{(\ell+1)\beta}{2}\right]
\end{equation}
where $(\ell+1)$ denotes the partner number with $\ell=0,1,2,...$,
and $\beta=\frac{\hbar^{2}\delta}{me^{2}}$ which is a
dimensionless quantity. This kind of superpotential choice leads
to the $(\ell+1)$-th member of the Hamiltonian hierarchy:
\begin{equation}
V_{\ell+1}(r)-E_{\ell+1}^{n=0}=W_{(\ell+1)}^{2}(r)-\frac{\hbar}
{\sqrt{2m}}\frac{d}{dr}W_{(\ell+1)}(r)~,
\end{equation}

\begin{equation}
V_{\ell+1}(r)=\frac{\hbar^{2}}{2m}\frac{\ell(\ell+1)
\delta^{2}e^{-2\delta r}}{(1-e^{-\delta r})^{2}}-e^{2}
\frac{\delta e^{-\delta r}}{(1-e^{-\delta
r})}\left[1-\ell(\ell+1)\frac{\beta}{2}\right]
\end{equation}
We introduce an expression for the bound-state energies of the 
above potential, considering the shape invariance requirement 
\cite{gendenshtein},
\begin{equation}
E_{\ell+1}^{n}=-\frac{me^{4}}{2\hbar^{2}}\left[\frac{1}
{(n+\ell+1)}- \frac{(n+\ell+1) (\beta)}{2}\right]^{2}~~;~~n=0,1,2,...
\end{equation}
and the corresponding ground-state eigenfunctions are
\begin{eqnarray}
\Psi_{(\ell+1)}^{n=0}(r)=N~exp\left(-\frac{\sqrt{2m}}
{\hbar}\int^{r}W_{(\ell+1)}(r')dr'\right)
\nonumber
\end{eqnarray}
\begin{equation}
=N~(1-e^{-\delta
r})^{\ell+1}exp\left\{-\frac{me^{2}}{\hbar^{2}}
\left[\frac{1}{(\ell+1)}-\frac{(\ell+1)\beta}{2}\right]r\right\}
\end{equation}
It is reminded that for a number of purposes it is convenient 
to have the wavefunction in such a compact analytical form. The 
first eigenfunction corresponds to the minimum energy for each 
$\ell$. In terms of the hierarchy of  Hamiltonians, we present 
here the lowest state wavefunctions for each member. 
The excited state wavefunctions can be determined [18,26] 
from the usual approach in SUSYQM.

For $\ell=0$ the potential in Eq. (4) leads to the usual 
Hulth\'{e}n potential which has an interesting property 
such that when the angular momentum is zero it is not shape 
invariant in the sense expressed in \cite{gendenshtein}. 
However, it is still possible to construct a general form 
of the potentials in the shape invariant super-family of 
Hamiltonians as seen in Eq. (4) where the first member 
corresponds to the Hulth\'{e}n potential which can be solved 
exactly in analytic form. One can easily verify that the 
energy eigenvalues and eigenfunctions for $\ell=0$ case of Eqs. (5)
and (6) are the same given in Refs. [5,6]. 
This supports the suggestion [17,22-24] that the
Gedenshtein's condition of shape invariance is sufficient 
but not a necessary condition  in the construction of 
exactly solvable but non-shape invariant potentials.

Eq.(4) can be rearranged as
\begin{equation}
V_{H}^{eff}(r)=V_{\ell+1}(r)=-e^{2}\frac{\delta e^{-\delta r}}{(1-e^{-\delta r})}
+\frac{\ell(\ell+1)\hbar^{2}\delta^{2}}{2m(1-e^{-\delta r)^{2}}}e^{-\delta r}
\end{equation}
which is known in literature as the approximate Hulth\'{e}n 
effective potential introduced by Greene and Aldrich 
\cite{greene} in their method to generate pseudo-Hulthen 
wave functions for $\ell\neq0$ states. For small $\delta r$, 
Eq. (7) is a good approximation to the realisticHulth\'{e}n 
effective potential, and unlike the original case the radial 
Schr\"{o}dinger equation for this potential is solvable
analytically through Eqs. (5) and (6). In addition, the partner 
potentials in Eq. (7) gives the necessary repulsive core due 
to angular momentum. For instance, for small $r$ the second 
term in Eq. (7) behaves as a $p$-wave centrifugal barrier for the
second member of the super-family. Since we know that the 
centrifigal potential is effective only in this region 
(i.e., small $r$), eigensolution of the potential for
$\ell=1$ in Eq. (6) can be regarded as the approximate 
$p$-wave solution for the Hulthen potential. Clearly, 
one can get other supersymmetric partners and their
solutions in an explicit form for $\ell\neq 0$ states. 
The present simple and elegant method is a clear cut of 
the iteration technique introduced by Laha {\it et al.}
[15,16]

For the sake of completeness, it is of interest to note 
that for small values of $\delta$, the potential in Eq. (7) 
closely approximates the effective Coulomb
potential rather well,
\begin{equation}
V_{H}^{eff}(r,\delta\cong 0)\rightarrow V_{C}^{eff}(r)=
-\frac{e^{2}}{r}+\frac{\ell(\ell+1)\hbar^{2}}{2mr^{2}}
\end{equation}
and the corresponding energy eigenvalue for the potential 
of Eq. (8), together with its ground state wavefunction 
for $\ell\neq 0$ states, obtained easily via Eqs. (5)
and (6) overlap with those, e.g. in Ref. \cite{dutt2}. 
This makes clear the work of Lam and Varshni \cite{lam} 
in which they showed that if one uses as trial functions
eigenvectors of theHulth\'{e}n potential rather than 
those of the simple Coulomb potential, excellent results 
for the energies of the  states of the screened Coulomb
potential can be obtained with simple variational wave functions 
containing only one parameter.

An important quantity of interest for the Hulth\'{e}n 
potential (and for other similar screened potentials) 
is the critical screening parameter $\delta_{c}$, which is that
value of $\delta$ for which the binding energy of the 
level in question becomes zero. Using Eq. (5), in atomic units,
\begin{equation}
\delta_{c}=\frac{2}{(n+\ell+1)^{2}}
\end{equation}
which works well for all $n$ values in case $\ell=0$ when 
compared to those in Table III of Ref. \cite{varshni}, 
but fails for non-zero angular momentum states. Consequently, 
the eigenenergies obtained via Eq. (5) for $\ell\neq 0$ 
states deviates from the accurate values obtained by 
numerical techniques and presented in Table I of Ref. 
\cite{varshni}. This may be understood as follow. 
If Eq. (7) is written in the form
\begin{equation}
V_{H}^{eff}(r)=-e^{2}\frac{\delta e^{-\delta r}}{(1-e^{-\delta r})}
+\frac{\ell(\ell+1)(\hbar^{2})}{2mr^{2}}+\left[\frac{\ell(\ell+1)\hbar^{2}\delta^{2}}
{2m(1-e^{-\delta r})^{2}}e^{-\delta r}-\frac{\ell(\ell+1)\hbar^{2}}{2mr^{2}}\right]
\end{equation}
the exact energy eigenvalues for the realistic effective 
Hulth\'{e}n potential may be given as
\begin{equation}
E_{H}^{n\ell}=-\frac{me^{4}}{2\hbar^{2}}\left[\frac{1}
{(n+\ell+1)}-\frac{(n+\ell+1)
\beta}{2}\right]^{2}+\Delta E
\end{equation}
where $\Delta E$ is the contribution, which does not 
appear in Eq. (9), due to the last term in Eq. (10). 
The clear interpretation of Eqs. (10) and (11) is that 
the potential barrier term prevents us to build the 
super-family as in the $\ell=0$ case, since the
potential-the first two terms in Eq. (10)-is not exactly 
solvable hence the supersymmetry is broken for $\ell\neq 0$ 
due to the potential barrier term. It is easy however 
to verify that for small values of $\delta$, $\Delta E$ goes 
to zero while Eq. (10) becomes an expression for the 
effective Coulomb potential in which case the accidental 
degenarcy is recovered as a natural consequence.

The usefulness of theHulth\'{e}n potential would be 
enhanced if one obtains an analytical expression for 
the exact energies of the non-zero angular momentum states.
The work along this line is in progress in the frame of 
broken supersymetry. Further, in the light of the 
supersymmetric solutions discussed in this letter we suggest here,
as an alternative to other various methods [4,7-14] 
investigating the bound-state properties of theHulth\'{e}n 
potential, an elegant approach for the calculation of the
whole energy spectrum of the potential using the first-order 
perturbation theory,
\begin{equation}
E_{H}^{n\ell}=E_{\ell+1}^{n}+\frac{\ell(\ell+1)\hbar^{2}}{2m}\int_{0}^{\infty}
\left[\psi_{\ell+1}^{n}(r)\right]^{2}\left(\frac{1}{r^{2}}-
\frac{\delta^{2}}{(1-e^{-\delta r})^{2}}e^{-\delta r}\right)dr
\end{equation}
which gives satisfactory values when compared (see Table 1) 
with the results obtained by the various methods for the 
eigenenergies of $\ell\neq 0$ levels. The accuracy of
the present calculations may be improved incorparating 
higher-order perturbations for in particular large values 
of the screening parameter.

\section{Conclusion}
We have obtained the exact analytical eigenfuntions and 
eigenvalues for the Hulth\'{e}n potential within the 
framework of SUSYQM for the case $\ell=0$. The approach 
consists of making an ansatz in the superpotential which 
satisfies the Riccati equation by an effective potential. 
For $\ell=0$ the effective potential obtained is identical 
to the Hulth\'{e}n potential. However, for $\ell\neq 0$ 
the effective supersymmetric potential has a slightly different 
structure than the Hulth\'{e}n potential. This deviation has
led us to introduce a simple expression that yields reasonable 
results for the non-zero angular momentum state energies. We 
stress that even though the problem has been attacked by 
different methods our simple and elegant methodology is powerful
because it provides an insight into the relation between 
theoretical partner Hamiltonians in the frame of supersymmetric 
quantum mechanics and physical states of the system considered. 
We hope to stimulate further examples of applications of SUSYQM
in important problems of  nuclear and atomic physics.

\hspace{1.0in}

\newpage\

\begin{table}
\begin{center}
\caption{Energy eigenvalues of the Hulth\'{e}n potential as a function of screening
parameter for various states in atomic units.}
\begin{small}
\begin{tabular}{ccccccccc}
\hline \\

$State$&$\delta$&Present&Numerical&Variational&Lai&Patil&Tang&Matthys\\

&&Calculations&Integration&(Ref. [4])&and Lin&(Ref. [9])&and Chan&and De Meyer\\

&&&(Ref. [4])&&(Ref. [7])&&(Ref. [12])&(Ref. [14])\\

\hline \\

2p &0.025 &0.1127605 &0.1127605 &0.1127605 &         &0.11276  &          &0.1127604\\

   &0.050 &0.1010425 &0.1010425 &0.1010425 &0.101043 &0.10104  &0.1010424 &0.1010425 \\

   &0.075 &0.0898478 &0.0898478 &0.0898478 &         &0.08985  &          &          \\

   &0.100 &0.0791794 &0.0791794 &0.0791794 &0.079179 &0.07918  &0.0791794 &0.0791794 \\

   &0.150 &0.0594415 &0.0594415 &0.0594415 &         &0.059445 &          &0.0594415 \\

   &0.200 &0.0418854 &0.0418860 &0.0418860 &0.041886 &0.041895 &0.0418857 &0.0418860 \\

   &0.250 &0.0266060 &0.0266111 &0.0266108 &         &         &          &          \\

   &0.300 &0.0137596 &0.0137900 &0.0137878 &0.013790 &         &          &0.0137900 \\

   &0.350 &0.0036146 &0.0037931 &0.0037734 &0.003779 &0.038375 &          &          \\
\\
3p &0.025 &0.0437068 &0.0437069 &0.0437069 &0.043707 &0.0437085&          &0.0437071\\

   &0.050 &0.0331632 &0.0331645 &0.0331645 &0.033165 &0.033185 &0.03316518&0.0331650 \\

   &0.075 &0.0239331 &0.0239397 &0.0239397 &         &0.0240165&          &          \\

   &0.100 &0.0160326 &0.0160537 &0.0160537 &0.016054 &0.01622  &0.01606772&0.0160537 \\

   &0.150 &0.0043599 &0.0044663 &0.0044660 &0.004466 &0.046995 &          &0.0044664 \\
\\
3d &0.025 &0.0436030 &0.0436030 &0.0436030 &0.043603 &0.0436025&          &0.0436030\\

   &0.050 &0.0327532 &0.0327532 &0.0327532 &0.032753 &0.032745 &0.0327532&0.0327532 \\

   &0.075 &0.0230306 &0.0230307 &0.0230307 &         &0.02299  &          &         \\

   &0.100 &0.00144832&0.0144842 &0.0144832 &0.014484 &0.01439  &0.0144842 &0.0144842 \\

   &0.150 &0.0132820 &0.0013966 &0.0013894 &0.001391 &0.0013755&          &0.0013965 \\
\\
4p &0.025 &0.0199480 &0.0199489 &0.0199489 &0.019949 &0.01995  &          &0.0199490\\

   &0.050 &0.0110430 &0.0110582 &0.0110582 &0.011058 &0.011075 &0.0110725 &0.0110583 \\

   &0.075 &0.0045385 &0.0046219 &0.0046219 &0.004622&0.0046585 &          &0.0046224 \\

   &0.100 &0.0004434 &0.0007550 &0.0007532 &0.000754 &0.000752 &          &          \\
\\
4d &0.025 &0.0198460 &0.0198462 &0.0198462 &0.019846 &0.019845 &          &0.0198462\\

   &0.050 &0.0106609 &0.0106674 &0.0106674 &0.010667 &0.01068  &0.0106690 &0.0106674 \\

   &0.075 &0.0037916 &0.0038345 &0.0038344 &0.003834 &0.003875 &          &0.0038346 \\
\\
4f &0.025 &0.0196911 &0.0196911 &0.0196911 &0.019691 &0.01969  &          &0.0196911\\

   &0.050 &0.0100618 &0.0100620 &0.0100620 &0.010062 &0.010045 &0.0100620 &0.0100619\\

   &0.075 &0.0025468 &0.0025563 &0.0025557 &0.002556 &0.002557 &          &0.0025563\\
\\
\hline \hline
\end{tabular}
\end{small}
\end{center}
\end{table}

\begin{table}
\begin{center}
\caption{Continue of Table 1.}
\begin{small}
\begin{tabular}{ccccccccc}
\hline \\

$State$&$\delta$&Present&Numerical&Variational&Lai&Patil&Tang&Matthys\\

&&Calculations&Integration&(Ref. [4])&and Lin&(Ref. [9])&and Chan&and De Meyer\\

&&&(Ref. [4])&&(Ref. [7])&&(Ref. [12])&(Ref. [14])\\
\hline \\

5p &0.025 &0.0094011 &0.0094036 &          &         &         &0.0094087 &        \\

   &0.050 &0.0026056 &0.0026490 &          &         &         &          &        \\
\\
5d &0.025 &0.0092977 &0.0093037 &          &         &         &0.0093050 &        \\

   &0.050 &0.0022044 &0.0023131 &          &         &         &          &        \\
\\
5f &0.025 &0.0091507 &0.0091521 &          &         &         &0.0091523 &        \\

   &0.050 &0.0017421 &0.0017835 &          &         &         &          &        \\
\\
5g &0.025 &0.0089465 &0.0089465 &          &         &         &0.0089465 &        \\

   &0.050 &0.0010664 &0.0010159 &          &         &         &          &        \\
\\
6p &0.025 &0.0041493 &0.0041548 &          &         &         &          &        \\
\\
6d &0.025 &0.0040452 &0.0040606 &          &         &         &          &        \\
\\
6f &0.025 &0.0038901 &0.0039168 &          &         &         &          &        \\
\\
6g &0.025 &0.0036943 &0.0037201 &          &         &         &          &        \\
\\
\hline \hline
\end{tabular}
\end{small}
\end{center}
\end{table}

\end{document}